\documentstyle[twoside,fleqn,npb,epsfig]{article}

\newcommand{\AmS}{{\protect\the\textfont2
  A\kern-.1667em\lower.5ex\hbox{M}\kern-.125emS}}

\title{NLO distributions for Higgs production at the LHC}

\author{J. Smith\address{C.N. Yang Institute for Theoretical Physics, 
 SUNY at Stony Brook, NY 11794-3840 USA}%
        \thanks{Work supported by the NSF Grant PHY-0098527.}}

\begin{document}

\begin{abstract}
We report on results for the NLO corrected differential distributions  
$d\sigma/dp_T$ and $d\sigma/dy$ for the process $p + p\rightarrow H + 'X'$, 
where $p_T$ and $y$ are the transverse momentum and rapidity of the
Higgs-boson $H$ respectively and $X$ denotes the inclusive hadronic state. 
All QCD partonic subprocesses have been 
included. The computation is carried out in the limit that the top-quark
mass $m_t \rightarrow \infty$.  Our calculations reveal that the dominant 
subprocess is given by $g + g \rightarrow H + 'X'$ but the reaction 
$g + q(\bar q) \rightarrow H + 'X'$ is not negligible.  
Also the $K$-factor representing the ratio between the next-to-leading 
order and leading order differential distributions varies from 
1.4 to 1.7 depending on the kinematic region and choice of parton densities. 
\end{abstract}

\maketitle

\section{Introduction}
The Higgs boson is the only particle in the standard model 
which has not yet been discovered. 
If the Higgs mass $m$ is between $110$ ${\rm GeV/c^2}$
and $200$ ${\rm GeV/c^2}$ then the dominant production mechanism at the LHC 
is $g + g \rightarrow H + 'X'$  where the Higgs boson couples to the gluons
via a top-quark loop. 
The leading order (LO) processes given by $g + g \rightarrow g + H$, 
$g + q(\bar q) \rightarrow q(\bar q) + H$ and 
$q + \bar q \rightarrow g + H$ were originally studied 
in \cite{hino}, \cite{ehsb} and \cite{kauff} from which one can derive
the transverse momentum ($p_T$) and rapidity ($y$) distributions of the
Higgs boson.  Fortunately the calculations simplify if one takes the
large top-quark mass limit $m_t \rightarrow \infty$.
In this case the triangle graphs are obtained from an effective Lagrangian 
describing the direct coupling of the Higgs boson
to the gluons, namely
\begin{eqnarray}
{\cal L}_{eff}=G\,\Phi(x)\,O(x) \,, 
\end{eqnarray}
where
\begin{eqnarray}
O(x)=-\frac{1}{4}\,G_{\mu\nu}^a(x)\,G^{a,\mu\nu}(x)\,.
\end{eqnarray}
Here $\Phi(x)$ represents the Higgs field and $G$ is an effective coupling
constant which is related to the Fermi coupling constant $G_F$ by
\begin{eqnarray}
\frac{G^2}{4\,\sqrt 2}&=&\left (\frac{\alpha_s(\mu_r^2)}{4\pi}\right )^2 G_F
\tau^2 F^2(\tau)
\nonumber \\ &&
\times {\cal C}^2\left (\alpha_s(\mu_r^2),\frac{\mu_r^2}{m_t^2}\right )\,,
\quad \tau=\frac{4m_t^2}{m^2}\,,
\end{eqnarray}
where $\alpha_s(\mu_r^2)$ is the running coupling constant which
depends on the renormalization scale $\mu_r$. The function 
\begin{eqnarray}
F(\tau)=1+(1-\tau) \arcsin^2 \frac{1}{\sqrt\tau}\,,
\end{eqnarray}
tends to $2/(3\tau)$ in the limit of large $\tau$. 
Further ${\cal C}$ is the coefficient function which originates from the
QCD corrections to the top-quark triangle graph describing the process
$H\rightarrow g + g$ in the limit $m_t \rightarrow \infty$. 
The lowest order contribution to $\cal C$ is available in \cite{gsz} 
and \cite{dawson}.

A comparison was made for the differential distributions 
of the LO processes in \cite{bagl} where it was shown 
that the large top-quark mass approximation is valid as long as $m$ 
and $p_T$ are smaller than $m_t$.
The NLO matrix elements for $g+g\rightarrow g+g+H$ etc., using the effective 
Lagrangian were computed in \cite{daka}, \cite{kdr} albeit in four dimensions. 
The one-loop virtual corrections to
the LO subprocesses were presented in \cite{schmidt},
where the computation of the loop integrals was performed in $n$-dimensions but
the matrix elements were still presented in four dimensions. 

\section{Method of calculation}

\begin{figure}[htb]
\vspace{4.9cm}
\includegraphics{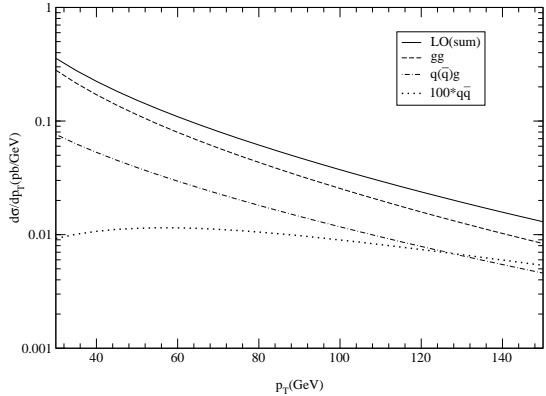}
\caption{
The differential cross section $d~\sigma/dp_T$ integrated over the whole 
rapidity range with $m=120~{\rm GeV/c^2}$ and $\mu^2=m^2+p_T^2$.
The LO plots are presented for the subprocesses $gg$ (long-dashed line),
$q(\bar q)g$ (dot-dashed line)
and  $100 \times (q\bar q)$ (dotted line)
using the parton density set MRST98(lo05a.dat).}
\label{fig:fig1}
\end{figure}

In this paper we present results from \cite{rsvn} where the NLO
corrections to the double differential distributions $d^2\sigma/dp_T/dy$
for Higgs boson production in hadron-hadron collisions were calculated.
Here we included all partonic subprocesses and used the g-g-H
coupling in Eq. (3).
A similar calculation has been performed in \cite{fgk} but our approach
differs from it in various aspects. First our calculation
is purely analytical and follows the calculation carried
out for the Drell-Yan process describing vector boson production in
hadron-hadron collisions (see \cite{arre}). The approach in \cite{fgk}
used helicity amplitudes and was based on the methods explained in \cite{fks}.
Also it was mainly numerical and has the advantage that it can
provide exclusive distributions.
In our calculation the matrix elements as well as the loop
integrals and phase space integrals are computed in $n$ dimensions.
Hence we could use results from \cite{pave}, \cite{been}, \cite{mmn} 
and \cite{bkns}. The advantage of this analytical approach is that one gets 
more insight into the structure of the radiative corrections. 
This is particularly important for the large corrections, due to soft 
gluon radiation and collinear fermion pair production, which arise
near the boundary of phase space, where the $p_T$ of the Higgs
boson gets large. Resummation of
this type of corrections has been carried out for the total cross section
in \cite{kls}. Resummation of small $p_T$ contributions due
to the Sudakov effect has been done in \cite{yuan}.
In view of the experimental problems to observe the Higgs boson,
a recalculation of all the NLO corrections is necessary to be
sure that the theoretical predictions are correct.
Finally we mention that another paper
has appeared on the NLO corrections to the $g + g \rightarrow H + g$ channel,
using the helicity framework \cite{glosser}.

For our computations the number of light flavours is taken
to be $n_f=5$ which holds for the running coupling,
the partonic cross sections and the number of quark flavour densities. 
Further we have chosen for our plots the parton densities obtained
from the sets MRST98 \cite{mrst98} CTEQ4 \cite{lai}, GRV98 \cite{grv}
and MRST99 \cite{mrst99}. For simplicity the factorization scale 
$\mu$ is set equal to the renormalization scale $\mu_r$.  For our plots 
we take $\mu^2=m^2+p_T^2$. Here we want to emphasize that the
magnitudes of the cross sections are extremely sensitive to the
choice of the renormalization scale because the effective g-g-H 
coupling constant $G\sim \alpha_s(\mu_r)$, 
which implies that $d\sigma^{\rm LO}\sim \alpha_s^3$ and
$d\sigma^{\rm NLO}\sim \alpha_s^4$. However the slopes
of the differential distributions are less sensitive to the 
scale choice if they are only plotted over a limited range.
For the computation of the g-g-H effective coupling constant in Eq. (3)
we chose the top-quark mass $m_t=173.4~{\rm GeV/c^2}$ and the Fermi constant
$G_F=1.16639~{\rm GeV}^{-2}=4541.68~{\rm pb}$.

\section{Results}

\begin{figure}[htb]
\vspace{4.9cm}
\includegraphics{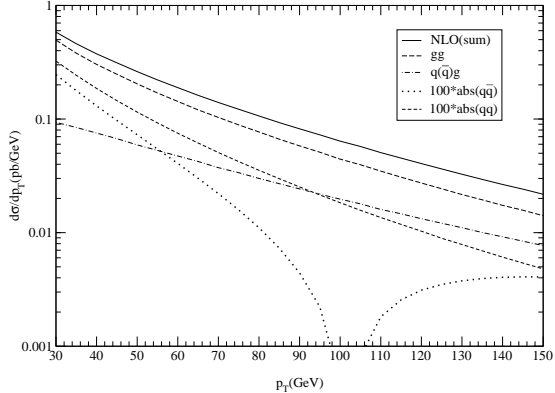}
\caption{
Same as Fig. 1 in NLO except for $100*{\rm abs}(q\bar q)$
(dotted line) and
the additional subprocess $100 \times {\rm abs}(qq)$ (short-dashed line)
using the parton density set MRST98(ft08a.dat). }
\label{fig:fig2}
\end{figure}

Here we will only give results for Higgs boson production in proton-proton 
collisions at the LHC center of mass energy $\sqrt S=14~{\rm TeV}$.  
Since the hadrons $H_1$ and $H_2$ are now identical the $y$ differential 
cross sections are symmetric (for results see \cite{rsvn}).
In order to compare with the results in \cite{fgk} we present LO and 
NLO differential cross sections in $p_T$, integrated over $y$, 
for $m=120$ GeV$/c^2$ and $\mu^2= m^2 + p_T^2$ in Figs. 1 and 2 respectively. 
The MRST98 parton densities \cite{mrst98} were used for these plots. 
We note that the NLO results from the $q\bar q$ and $qq$ channels 
are negative at small
$p_T$ so we have plotted their absolute values multiplied by 100.
It is clear that the $gg$ subprocess dominates but the $q(\bar q)g$-subprocess
is also important, yielding one-third of the total at large $p_T$. The 
LO(sum) and NLO(sum) are slightly lower when $m_t$ is taken very large 
in which case we agree with the results in \cite{fgk}.

There are several uncertainties which affect the predictive power of the
theoretical cross sections. The first one concerns scale dependence.
In the case of the $p_T$-distribution one observes a small
reduction in the scale dependence while going from LO to NLO. This
reduction becomes more visible when we plot the quantity
\begin{eqnarray}
N\left (p_T,\frac{\mu}{\mu_0}\right )
=\frac{d\sigma(p_T,\mu)/dp_T}{d\sigma(p_T,\mu_0)/dp_T}
\end{eqnarray}
in the range $0.1 < \mu/\mu_0 < 10$ at fixed values of $p_T=$ 30, 70 and
~{\rm 100 GeV/c}, see Fig.3.
The upper set of curves at small $\mu/\mu_0$ are for LO and the
lower set are for NLO. Notice that the NLO plots at 70 and 100 are
extremely close to each other and it is hard to distinguish between them.
Further one sees that the slopes of the LO curves are
larger that the slopes of the NLO curves. This is an indication that
there is better stability in NLO, which was expected.
However there is no sign of a flattening or an optimum in either of these
curves which implies that one will have to calculate the
differential cross sections in NNLO to find a better stability under
scale variations.

\begin{figure}[htb]
\vspace{5.5cm}
\includegraphics{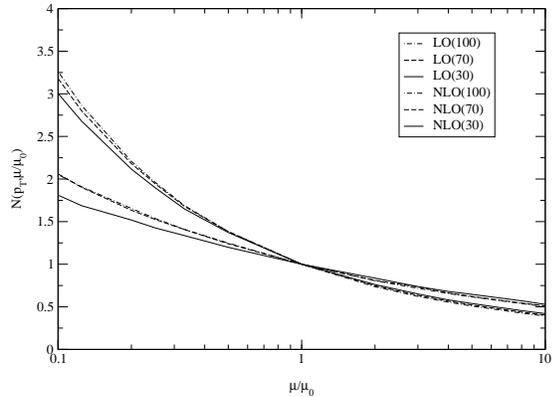}
\caption{
The quantity $N(p_T,\mu/\mu_0)$ plotted in the range
$0.1<\mu/\mu_0<10$ at fixed values of $p_T$
with $m=120~{\rm GeV/c^2}$ and $\mu_0^2=m^2+p_T^2$
using the MRST98 parton density sets.  The results are shown for
$p_T=30~{\rm GeV/c}$ (solid line), $p_T=70~{\rm GeV/c}$ (dashed line),
$p_T=100~{\rm GeV/c}$ (dot-dashed line).
The upper three curves on the left hand side are the LO results whereas
the lower three curves refer to NLO.}
\label{fig:fig3}
\end{figure}

\begin{figure}[htb]
\vspace{5.5cm}
\includegraphics{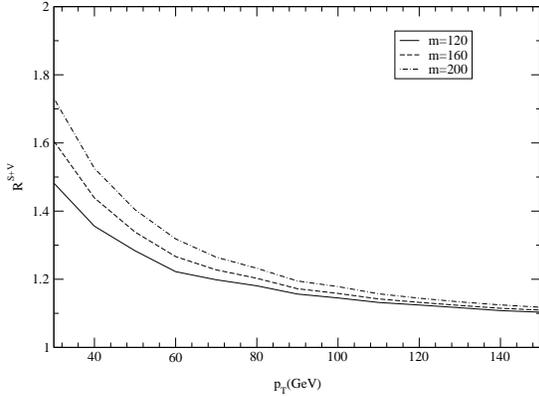}
\caption{
The ratio $R^{\rm S+V}$ for the $p_T$
distributions using the set MRST99 with $\mu^2=m^2+p_{T,{\rm min}}^2$
and various Higgs masses given by
$m=120~{\rm GeV/c^2}$ (solid line),
$m=160~{\rm GeV/c^2}$ (dashed line) and
$m=200~{\rm GeV/c^2}$ (dot-dashed line).}
\label{fig:fig4}
\end{figure}

The second uncertainty concerns the rate of convergence
of the perturbation series which is indicated by the $K$-factor defined by
$K=d~\sigma^{\rm NLO}/d~\sigma^{\rm LO}$ 
and finally there is the dependence of the cross section on the specific
choice of parton densities, which can be expressed by the factors
$R^{\rm CTEQ}=d~\sigma^{\rm CTEQ}/d~\sigma^{\rm MRST}$ and 
$R^{\rm GRV}=d~\sigma^{\rm GRV}/d~\sigma^{\rm MRST}$.
The above quantities were studied in our paper, where it was shown  
that there is still a large uncertainty in our predictions because the
$K$-factor varies from approximately 1.4 to 1.7 depending on the parton
density set.  The latter uncertainty is mainly due to the
small $x$ behaviour of the various gluon densities since
both the partonic cross sections and the gluon densities increase very
steeply at decreasing $x$. Finally we have defined the 
soft-plus-virtual (S+V) approximation and computed the ratio
$R^{\rm S+V}=d\sigma^{\rm S+V}/d\sigma^{\rm EXACT}$. This
approximation is quite reasonable (see Fig. 4) provided
$p_{T,{\rm min}}>100~{\rm GeV/c}$ in spite of the fact that $x$ is still
too small to belong to the large $x$-region. This means that this
approximation can be used to resum the large corrections due to
S+V gluons in order to obtain a better estimate of the all order 
corrected cross section.

\end{document}